\documentclass[12pt,cite,epsf,epsfig,psfrag]{article}
\usepackage{epsfig}
\usepackage{amsmath, graphics, setspace}
\usepackage{graphicx}
\usepackage{psfrag}
\usepackage{amssymb}
\usepackage{subfigure}

\newcommand {\Tr}{\mbox{Tr}}
\newcommand{\be}{\begin{equation}}
\newcommand{\ee}{\end{equation}}
\newcommand{\bea}{\begin{eqnarray}}
\newcommand{\eea}{\end{eqnarray}}

\newcommand{\nn}{\nonumber}
\begin{document}
\baselineskip=0.6cm
\begin{titlepage}
\begin{center}
\vskip .2in
{\Large \bf Correlators of Giant Gravitons from dual ABJ(M) Theory}
\vskip .5in
{\bf Shankhadeep Chakrabortty$^*$\footnote{e-mail: sankha@iopb.res.in} and Tanay K. Dey$^{\ddagger}$\footnote{e-mail: tanay.dey@nucleares.unam.mx}}\\
\vskip .25in
{\em $^*$Institute of Physics,\\
Bhubaneswar 751 005, INDIA}
\vskip .2in
{\em $^{\ddagger}$Departamento de F$\acute{i}$sica de Altas Energ$\acute{i}$as, \\
Instituto de Ciencias Nucleares,\\
Universidad Nacional Aut$\acute{o}$noma de M$\acute{e}$xico,\\
Apartado Postal 70-543, M$\acute{e}$xico D.F. 04510, M$\acute{e}$xico}
\end{center}
\vskip .3in
\begin{center}
{\bf ABSTRACT}
\end{center}

\begin{quotation}\noindent
\baselineskip 18pt
We generalize the operators of ABJM theory, given by Schur polynomials, in ABJ theory by computing the two point functions in
the free field and at finite $(N_1,N_2)$ limits. These polynomials are then identified with the states of the dual gravity theory.
 Further, we compute correlators among giant gravitons as well as between giant gravitons and ordinary gravitons through the
 corresponding correlators of ABJ(M) theory. Finally, we consider a particular non-trivial background produced by an operator with
 an $\cal R$-charge of $O(N^2)$ and find, in presence of this background, due to the contribution of the non-planar corrections,
the large $(N_1,N_2)$ expansion is replaced by $1/(N_1+M)$ and $1/(N_2+M)$ respectively.
\end{quotation}
\vskip .3in
\end{titlepage}
\vfill
\eject

\section{Introduction}

\noindent
Strongly coupled gravity is generally very difficult to study. Perturbative analysis fails and there are no systematic approaches to
study gravity. Consequently questions related to transitions among brane like objects called giant
gravitons\cite{Herzog,Susskind1,Grisar,Hashimoto}, or even between giant gravitons and small ordinary gravitons become difficult
 to address within conventional frame work. However,
the AdS/CFT correspondence\cite{Malda,Gub,Wit,Ahar} offers a systematic way to study these transitions via proper
identifications between the states in the gravity theory and the tractable relevant operators of the dual gauge theory.
According to this conjecture, the gravitons of the gravity theory are identified with the single trace operators of the
${\cal N}=4$, $SU(N)$ super Yang-Mills gauge theory having $\cal R$-charge of the order one\cite{Wit} and initially,
the giant gravitons which are only wrapped up in the $S^5$ component of the $AdS_5 \times S^5$ geometry are identified with
the sub-determinant operators having $\cal R$-charge of the order $N$\cite{Bala}. Subsequently, the giant gravitons expanded inside any one
of the $S^3$ of $ AdS_5$ or $S^5$ components of the same geometry are holographically mapped to the operators known as Schur
polynomials\cite{Sanjay,Dberen,LLM}. All these operators are built out of the fields in the Yang Mills supermultiplet and are labeled by
Young diagrams with boxes equal to the $\cal R$-charge of the operators. In particular, the single giant graviton wrapped
 inside the $S^5$ component is mapped into the single column Young diagram with maximum $N$ number of rows, whereas the
same wrapped inside the $AdS_5$ component can be associated with the single row Young diagram without any bound on the number
of columns. The restricted Schur polynomials describe the excited giant gravitons\cite{Vbala,Dmello1,Dmello2,Dmello3,Dmello4}.\\

\noindent
The gauge theory computations\cite{Bala,Sanjay}, dual to the  non perturbative bulk transitions among giant
gravitons as well as transitions between giant gravitons and small gravitons are based on two normalization prescriptions:
one is called overlap of state normalization and the other one is called multi-particle normalization.
The first normalization scheme gives rise to the transition amplitudes with magnitude less than or equal to one and hence has a
natural probabilistic interpretation. The second one gives rise to the amplitudes that have unbounded growth in $N$, spoiling a naive
probabilistic interpretation. The problem is resolved in \cite{Brown}, using the full space-time dependence of the correlators.
Due to the presence of space-time dependence, the normalization factors of
the correlators on manifolds of non-trivial topologies have brought the importance of the CFT factorization equations,
directly into the picture. CFT factorization relates appropriate normalization with the genus number of the manifold.\\

\noindent
Recently, as a new candidate for AdS/CFT duality the so-called ABJM model named after Aharony, Bergman,
Jafferis and Maldacena\cite{Abjm}(see also \cite{Benna,Bhat,Bandres} for further results) is discovered.
Recognising that this theory may help us understand physics of multiple $M2$ branes, several investigations have been
carried out to analyze ABJM theory. In this model duality holds between the M theory on
$AdS_4 \times S^7/{\mathbb {Z}}_k$ and ${\cal N} = 6$ super Chern-Simons theory on the 3-dimensional boundary
of the AdS space. The ABJM model is deemed to be the world volume theory of $N$ number of $M2$-branes stack at the
singularity of the orbifold $\mathbb{C}^4/\mathbb{Z}_k$. The parameter $k$ denotes the Chern-Simons level. For large $k$, the
11-dimensional supergravity reduces to 10-dimensional IIA super string theory on $AdS_4 \times \mathbb{CP}^3$.
The gauge group of this three dimensional Chern-Simons matter theory is $U(N)_k \times U(N)_{-k}$. The field content of the theory
includes four complex scalar fields $(A_1, A_2, B_1^\dagger, B_2^\dagger)$ with conformal dimension and $\cal R$-charge both equal
 to $1/2$.
These fields are $N \times N$ matrices transforming in the bi-fundamental representation of the gauge group. This theory enjoys a sensible
large $N$ expansion like ${\cal N}= 4$ SYM theory where the large $N$ expansion parameter is $1/N$.\\

\noindent
Immediately after, as a further generalization, another model is proposed by Aharony, Bergman,
 Jafferis\cite{Abj}) and named as ABJ model, by modifying, ABJM gauge group
$U(N)_k \times U(N)_{-k}$ to $U(N_2)_k \times U(N_1)_{-k}$ for the same Chern-Simons matter fields, with $N_2 \ge N_1$.
The matter fields here are given by a rectangular matrices of dimension $N_1 \times N_2$. In the gravity dual,
 the brane construction is equivalent to the low energy theory of $(N_2-N_1)$ fractional M2-branes sitting at
 the $\mathbb{C}^4/\mathbb{Z}_k$ singularity in one sector and $N_1$ M2 branes freely moving around on the other.
The geometric structure of the gravity remains the same as in ABJM theory, with only an additional discrete holonomy of the M
 theory 3-form field on a torsion 3-cycle in $S^7/\mathbb{Z}_k$ and a two form $B_{NS}$ holonomy turned on over
$\mathbb{CP}^1\subset\mathbb{CP}^3$. On the field theory side, this generalization is classically straightforward
and still enjoys ${\cal N}=6$ superconformal symmetry. However the quantum consistency of the theory requires the bound
 $|N_2 - N_1| \le |k|$ and unlike ABJM theory, this theory has two large $N$ expansion parameters $1/N_1$ and $1/N_2$.\\

\noindent
The study of M theory or their $IIA$ descendants from the dual free field regime has been initiated in
 \cite{Beren,Sheikh,Beren1} by constructing the half-BPS operators for small $\cal R$-charge. Like the ${\cal N}=4$ SYM theory,
these operators are also labeled by Young tableaux. The number of boxes in the Young tableaux are equal to the $\cal R$-charge
 of the corresponding operator. These half-BPS operators are well described by single trace operators and are to be identified
with the giant gravitons of the dual gravity theory constructed in
\cite{Nishioka,Hamilton,Hamilton1,Murugan,Lozano,Murugan1}. In $AdS_4 \times S^7/\mathbb{Z}_k$, these giant gravitons are either
 spherical M2-brane growing in the $AdS_4$ or M5-brane wrapping an $S^5 \subset S^7$. Whereas for $AdS_4 \times \mathbb{CP}^3$,
the M2-brane is replaced by a D2-brane growing into $AdS_4$ and instead of M5-brane, D4-brane wrapping on some circle of
$\mathbb{CP}^3$. The single giant graviton nested in $AdS_4$ is described by an operator labeled by a Young
tableau of one row with unrestricted number of columns and the same wrapped on  $\mathbb{CP}^3$ is described by a Young tableau
which has one column and maximum number of rows given by the smallest of $N_1,N_2$.\\

\noindent
In \cite{Dey}, it is shown that, for the ABJM model, if the $\cal R$-charge of the gauge operator is greater than $N^{2/3}$,
the role of single trace operators is replaced by Schur polynomials. Although in the current setting, the Schur polynomials
do not provide a complete basis to study the gauge theory, they diagonalize the two point function in the free field limit. These operators
are very useful to study the large $N$ expansion associated with the theory for both trivial and non-trivial backgrounds.
 For non-trivial background produced by an operator with  $\cal R$-charge of the order of $N^2$, large $N$ expansion becomes $1/(N+M)$,
 where $M$ is the number of columns in the representing Young diagram of the operator\cite{Dmello,Koch4,dey}
and is of the order $N$. It is natural to expect the same phenomena will be repeated for the ABJ theory also.
This is what we aim to demonstrate in this paper. We first extend our study to construct the correct gauge
invariant operator in the ABJ theory by generalizing the Schur polynomial constructed for ABJM theory.
We then find out the the realization of the duality between giant gravitons and the Schur polynomials as the gauge invariant
operators for both ABJ(M) theory. Further, we study the transition probabilities among giant gravitons as well as between giant
gravitons and ordinary gravitons by analyzing the corresponding gauge theory correlators ( constrained to the appropriate normalization
prescribed in \cite{Brown} ). Finally, we work out the modification of the large $N$ expansions in presence of the non-trivial background.\\

\noindent
This paper is organized as follows: in section 2, we generalize the Schur polynomials of ABJM theory for ABJ
theory and also identify them with the gravitons and giant gravitons of the dual gravity theory. With the aim of
finding out the transition probability among giant gravitons and between giants and gravitons we discuss CFT
factorization and its interpretation as probability in section 3. Appropriate transition probability needs to know
the genus number of the manifold. In section 4, 5 and 6 we compute probability on genus zero, one and higher genus among the different states of the gravity and in section 7 we enumerate the same results for ABJM theory. We also study the large $N$ expansions in non-trivial background in section 8.
Finally, we conclude in section 9.

\section{Schur Polynomial}

\noindent
In this section we generalize the Schur polynomial of the ABJM theory to the one for the ABJ theory. With this goal in mind,
we closely follow the derivation of the Schur polynomial of ABJM theory keeping track of whether
 the gauge indices are associated with the gauge groups, $U(N_1)$ or $U(N_2)$. Depending on the choice of the gauge group,
 contraction of the gauge indices gives rise to the factors associated with either $N_1$ or $N_2$. Following the logic as described in
 \cite{Dey}, we can write down the simplest gauge invariant half-BPS operator as $\prod_{n_i}[\Tr((AB^\dagger)^l)]^{n_i}.$
We consider the compact notation A and B to depict the four complex scalars $A_1,A_2$ and $B_1,B_2$ respectively.
Since these complex scalar fields transform under the bi-fundamental representation of the gauge group, therefore in
the matrix notation, we can write $A$ and $B^\dagger$ in the following way $$A_j^i \;\; {\rm and} \;\; (B^\dagger)_i^j.$$
Unlike ABJM now $i$ and $j$ are gauge indices of  $U(N_1)$ and $U(N_2)$ respectively. Here $l$ counts the amount of $\cal R$-charge
of the operator. According to \cite{Beren,Sheikh}, these operators are represented by Young tableaux consisting of boxes equal
 to the number of $(AB^\dagger)$ fields and at most the smallest of $(N_1,N_2)$ rows.  In \cite{Dey} we have shown that single
 trace operators are not valid basis to study the ABJM gauge theory in the large $\cal R$-charge limit. The correct description
of the gauge theory operators are Schur polynomials. Therefore to check the validity of the single trace operator in this theory, following \cite{Bala,Dey}, we compute the
 3-point function of two different operators. In the leading order the result is
\begin{equation}
\frac{\Big\langle{\cal{O}}_1{\cal{O}}_2^\dagger\Big\rangle}{\sqrt{\Big\langle{\cal{O}}_1
{\cal{O}}_1^\dagger\Big\rangle}\sqrt{\Big\langle{\cal{O}}_2
{\cal{O}}_2^\dagger\Big\rangle}} \sim \frac{\sqrt{l l_1 l_2}}{N_1}+ \frac{\sqrt{l l_1 l_2}}{N_2}.
\end{equation}
Where
\begin{eqnarray}
\nonumber {\cal{O}}_1 &=& \Tr\big((AB^\dagger)^l\big)\\\nonumber
{\rm and}\quad {\cal{O}}_2 &=& \Tr\big((AB^\dagger)^{l_1}\big)\Tr\big((AB^\dagger)^{l_2}\big) \hspace{.4in}{\rm with}\hspace{.4in}l_1+l_2=l.
\end{eqnarray}
To compute this three point function we use the basic Wick contraction between two fields. One simple example of this contraction is
$$\Big\langle \Tr\big(AB^\dagger\big)\Tr\big(A^\dagger B\big)\Big\rangle = \Big\langle A_{j_1}^{i_1}{B^\dagger}_{i_1}^{j_1}
{A^\dagger}_{i_2}^{j_2}B_{j_2}^{i_2}\Big\rangle =  \delta_{i_2}^{i_1}\delta_{j_1}^{j_2}\delta_{j_2}^{j_1}\delta_{i_1}^{i_2}= N_1 N_2.$$
Since $i$ and $j$ are the gauge indices of $U(N_1)$ and $U(N_2)$ group respectively, the sum over $i$ gives $N_1$ and the same over $j$
gives $N_2$. However, we drop the space-time dependence in this computation, we can easily bring them back when needed.
 The result of the three point function says that at large $(N_1,N_2)$ it vanishes if the factor $\sqrt{ll_1l_2}$ is less
than $(N_1,N_2)$ and diverges otherwise. Therefore like ABJM model, in this case also, the trace operator will be the
 gauge invariant operator if $\cal R$-charge $l$ is less than the smaller one between $N_1^{2/3}$ or $N_2^{2/3}$.
 However, we should remember that we have just computed one specific three point function and not the most general correlator of
the theory. The BMN type analysis in ${\cal N} = 4$ super Yang-Mills theory as well as ABJM theory is more exhaustive to conclude
about the limit of the $\cal R$-charge\cite{BMN,TT}. These study suggest that it must be less than $\sqrt N$ to suppress the
 non-planar corrections those are important even before $\cal R$-charge gets to $N^{2/3}$. Thus we rather say that if $\cal R$-charge
 of the operator in the ABJ(M) theory is greater than the smallest of $(\sqrt{N_1},\sqrt{N_2}) $, for the correct description, we need
 a new basis and our natural choice is Schur polynomial given by
\begin{equation}
\chi_R(AB^\dagger)=\frac{1}{n!}\sum_{\sigma\in S_n}\chi_R(\sigma)\Tr\big(\sigma(AB^\dagger)\big)
\end{equation}
 where $R$ is the representation of a specific Young diagram with $n$ boxes. This Young diagram labels both the representation
of unitary gauge group and symmetric group $S_n$. $\chi_R(\sigma)$ is the character or trace of the element $\sigma \in S_n$
in the representation $R$.\\

\noindent
The two point function for this theory can  easily be calculated using the same procedure of \cite{Sanjay,Dey}.
 The result of the two point function of our interest is
\begin{equation}
\bigg\langle\chi_R(AB^{\dagger})\chi_S(A^{\dagger}B)\bigg\rangle = \Big( n!\frac{Dim_{N_1}(S)}{d_S}\Big)\Big
( n!\frac{Dim_{N_2}(S)}{d_S}\Big)\delta_{RS} = f_S^{N_1}f_S^{N_2}\delta_{RS}.
\label{2point}
\end{equation}
Here $R$ and $S$ represent the Young diagrams with $n$ number of boxes for symmetric group $S_n$.  $d_S$ is the dimension
of a representation $S$ of the permutation group $S_n$. $Dim_{N_1}(S)$ and $ Dim_{N_2}(S)$ are the dimensions of the
representation of the unitary group $U(N_1)$ and $U(N_2)$ respectively. $f_S^{N_1}$ and $f_S^{N_2}$ are the product of
the weights of the same Young diagram  but for the gauge group $U(N_1)$ and $U(N_2)$ respectively. In this calculation we
 suppress the space-time dependence. The presence of delta function says that Schur polynomials satisfy the orthogonality
 condition in the free field limit and therefore these Schur polynomials are the correct gauge invariant operators to study
 the ABJ(M) theory for large $\cal R$-charge. Recall that for usual Schurs in ${\cal N}=4$ super Yang-Mills theory,
 if we have more than $N$ boxes in a column the product of weights vanishes\cite{Sanjay}; this is the stringy exclusion principle of
 $AdS_5$\cite{Strominger}. Equation(\ref{2point}) shows a similar behavior for ABJ(M) theory. When the number of boxes
 is more than the smallest of $(N_1,N_2)$ the two point correlator and, hence, the operator vanish. Thus it is the smallest
 of $(N_1,N_2)$ which sets the bound on the stringy exclusion principle.\\

\noindent
In order to bring back the space-time dependence of the two point function we first find out the Green's function $G(x-y)$ for
 three dimensional gauge theory. This Green's function is the solution of the differential equation
\be
\Delta_xG(x-y)= -\delta^3(x-y).
\ee
The solution is given by
\be
G(x-y)=\frac{1}{4\pi|x-y|}.
\ee
Therefore the above two point function, with space-time dependence included, takes the form
\begin{equation}
\bigg\langle\chi_R(AB^{\dagger})(x)\chi_S(A^{\dagger}B)(y)\bigg\rangle = \frac{f_S^{N_1}f_S^{N_2}\delta_{RS}}{\big(4\pi|x-y|\big)^{2\Delta_R}}
\end{equation}
where $\Delta_R$ is the conformal dimension of the operator $\chi_R(AB^{\dagger})$.\\

\noindent
The three point and multi point functions can easily be calculated from this two point function by using the product rule of Schur polynomials. The product rule of Schur polynomials is as follows
\begin{equation}
\chi_{R_1}(AB^\dagger)\chi_{R_2}(AB^\dagger)=\sum_S g(R_1,R_2;S)\chi_{S}(AB^\dagger).
\label{product1}
\end{equation}
Here the Littlewood-Richardson number $g(R_1,R_2;S)$ counts the number of times irreducible representation $S$ appears in the direct product of irreducible representations $R_1$ and $R_2$. By repeated use of this product rule we can write the direct product of
$\chi_{R_1}(AB^\dagger)\chi_{R_2}(AB^\dagger)\cdots\chi_{R_l}(AB^\dagger)$ as \cite{Dmello1}
\begin{eqnarray}
\nonumber\prod_{i=1}^l\chi_{R_i}(AB^\dagger) &=& \hspace{-.2in}\sum_{S_1,S_2\cdots S_{l-2},S} \hspace{-.05in}
g(R_1,R_2;S_1)g(S_1,R_3;S_2)\cdots g(S_{l-2},R_l;S)\chi_{S}(AB^\dagger)\\
&=&\hspace{.05in}\sum_S g(R_1,R_2\cdots R_l;S)\chi_{S}(AB^\dagger).
\label{product2}
\end{eqnarray}
By using this product rule we can have the three point function as
\be
\Bigg\langle\chi_{R_1}(AB^{\dagger})(x_1)\chi_{R_2}(AB^{\dagger})(x_2)\chi_S(A^{\dagger}B)(y)\Bigg\rangle
=\frac{g(R_1,R_2;S)f_S^{N_1}f_S^{N_2}}{\{4\pi(y-x_1)\}^{2\Delta_{R_1}}\{4\pi(y-x_2)\}^{2\Delta_{R_2}}}
\ee
where $\Delta_{R_1}$ and $\Delta_{R_2}$ are conformal dimensions of the operator $\chi_{R_1}(AB^{\dagger})$ and $\chi_{R_2}(AB^{\dagger})$. Combination of these two conformal dimensions gives the conformal dimension of $\chi_S(A^{\dagger}B)$. Similarly we can find out the multi point function as
\bea
&&\hspace{-.45in}\Bigg\langle\hspace{-.04in}\chi_{R_1}(AB^{\dagger})(x_1)\chi_{R_2}(AB^{\dagger})(x_2)\cdot\cdot \chi_{R_l}(AB^{\dagger})(x_l) \chi_{S_1}(A^{\dagger}B)(y)\chi_{S_2}(A^{\dagger}B)(y)\cdot\cdot\chi_{S_k}(A^{\dagger}B)(y) \hspace{-.04in}\Bigg\rangle\nn\\&=&\sum_S \frac{g(R_1,R_2\cdots R_l;S)\,f_S^{N_1}f_S^{N_2}\;  g(S_1,S_2\cdots S_k;S)}{\{4\pi(y-x_1)\}^{2\Delta_{R_1}}\cdots\{4\pi(y-x_l)\}^{2\Delta_{R_l}}}.
\eea
The class of correlators that we study here are the analog of the extremal correlators of the ${\cal N}=4$ super Yang-Mills theory.
Motivated by our experience with ${\cal N}=4$ super Yang-Mills theory, we make a guess that these correlators are the exact answer
i.e. they will not receive any higher loop corrections in the 'tHooft coupling. We need to compute the Feynman graphs to confirm
the guess. As before, $\Delta_{R_1}\cdots \Delta_{R_l}$ are the conformal dimensions of
$ \chi_{R_1}(AB^{\dagger})\cdots\chi_{R_l}(AB^{\dagger})$
and the sum of these conformal dimensions gives the total conformal dimension of the representation $S$.\\

\noindent
According to \cite{Beren,Sheikh,Beren1} trace operators are identified with the giant gravitons of the dual gravity theory and
the $\cal R$-charge of the operators with the angular momentum of the giant gravitons. The trace operator represented by a single
row Young diagram is mapped into the giant graviton which grows within the $AdS$ part of the geometry and is called as AdS giant and the
operator corresponding to Young diagram of single column with maximum number of boxes equal to the smallest of $(N_1,N_2)$ is mapped
into the giant graviton which wraps in $S^7$ or $\mathbb{CP}^3$, and is called as sphere giant. Similarly we can also map this Schur polynomial with the giant
 graviton of the dual gravity theory. The Schur polynomial associated with Young diagram of single column i.e. fully antisymmetric
representation of symmetric group $S_n$ with at most the smallest of $(N_1,N_2)$ number of boxes is mapped into a sphere giant and the
operator represented by single row Young diagram i.e. fully symmetric representation of the same group corresponds to an AdS giant.
In this paper following\cite{Brown}, we use the notation $\chi_{[L]}$ for the AdS giant and $\chi_{[1^{L}]}$ for the sphere giant with angular
momentum of $L$ units. If the sphere giant brane wraps $k$ times, the corresponding Schur polynomial will have Young diagram of $k$
 number of columns whereas the number of boxes associated with each of them is of the same order as the smaller one between $N_1$ and $N_2$.
For the $k$ number of wrapping of  AdS giant
 within the circle of $AdS$ the corresponding Young diagram will have $k$ number of rows where the maximum value of $k$ can be the
smallest of $(N_1,N_2)$. Schur polynomial represented by Young Diagram $R$ associated with small $\cal R$-charge, i.e conformal dimension
 $\Delta_R = O(1)$ is associated with KK state of the gravity. They can be written as sums of products of small numbers of traces.\\

\noindent
By using this mapping we can compute the gravity correlators between KK states, giant gravitons and among KK states and giant gravitons
 from the corresponding  gauge theory correlators. These correlators give the probability for the state created by the operator at
 a particular point of the space-time to evolve into the state created by another operator at different point of the space-time by
 proper normalization. Therefore, it is very crucial to normalize the gauge theory correlators in a consistent way to get the
 probability less than 1. In the literature, there exist two types of normalization\cite{Brown},
namely the overlap of state normalization and the multi particle normalization. Both prescriptions consist of two parts,
 one depends on gauge indices and other is a function of space-time coordinates of involved operators. Without the space-time
dependence, the first scheme gives the probabilities within one but the second procedure suffers from growth in $N$. However by
including the space-time part, the problem is resolved in \cite{Brown}. So to figure out the normalization factor, we need to know
the topology of the space on which the operators live. In conformally invariant field theories, the analysis of correlators on different
 topologies and the relation between them is generically known as CFT factorization leading to factorization equations and inequalities
 in the specific dimension. In the next section we briefly review the CFT factorization for different topologies and interpret a
normalized version of this as the probability.

\section{From CFT factorization to probability interpretation}
\begin{figure}[h]
\begin{center}
\begin{psfrags}
\psfrag{A}[][]{$\hspace{.3cm} =\sum_B$}
\psfrag{B}[][]{$\hspace{.1cm}\frac{|B\rangle\langle B|}{\langle B|B\rangle}$}
\includegraphics[width=13cm]{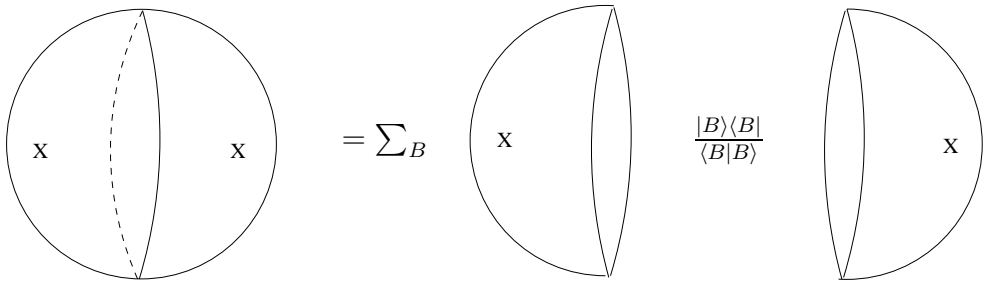}
\caption{The operator insertions are represented by cross marks.}
\label{sphere}
\end{psfrags}
\end{center}
\end{figure}

\noindent
This section generalizes the discussion of \cite{Brown} to ABJ(M) theory.
Factorization can be explained in the following intuitive way. We consider a manifold $M$ with two operator insertions and compute the corresponding correlator $Z_M$. Now we cut the manifold $M$ along a boundary $\mathbb{B}$ into two constituent manifolds, $M_1$, $M_2$ with one operator insertion in each and compute the correlators $Z_{M_{1}}$ and $Z_{M_{2}}$ accordingly, constrained to the fact that all possible boundary configuration should be taken into account. The CFT factorization suggests,
\begin{equation}
Z_M = \sum_{B} Z_{M_{1}}(B) \times Z_{M_{2}}(B).
\end{equation}
In the context of overlap state normalization we consider a manifold $S^3$ with two operator insertions.
Then we cut it into two manifolds with one boundary $\mathbb{B}$ having one operator in each as depicted in fig \ref{sphere}.
Now the factorization in conformal field theory relates this n-point correlator to lower point correlator as
\be
\Big\langle{{\cal O}^{\dag}}({x^{*}}){\cal O}(x)\Big\rangle = \sum_{B} \frac{\big\langle {{\cal O}^{\dag}}({x^{*}})B(y)\big \rangle\big\langle B^{\dag}
(y^{*}){\cal O}(x)\big \rangle}{\big\langle B^{\dag}
(y^{*})B(y)\big\rangle}
\label{overstatnorm}
\ee
where $\cal O'$s and  $B'$s are the local operators defined on the manifold of  interest and on the boundary cut respectively.
While defining equation({\ref{overstatnorm}}) we assume that we are working in a basis which diagonalizes the metric on the space of
local operators and the conjugation operation is executed by reversing the Euclidean time coordinate. This can be generalized for the
extremal correlators where the operators are localized at number of different points:
\bea
&&\Big\langle{{\cal O}^{\dag}}_1({x^{*}}_1){{\cal O}^{\dag}}_2({x^{*}}_2)\cdots
{{\cal O}^{\dag}}_k({x^{*}}_k)
{\cal O}_1(x_1){\cal O}_2(x_2)\cdots {\cal O}_k(x_k)\Big\rangle =  \nonumber  \\
&& \hspace{-.3in}\sum_{B} \frac{\Big\langle {{\cal O}^{\dag}}_1({x^{*}}_1){{\cal O}^{\dag}}_2
({x^{*}}_2)\cdots {{\cal O}^{\dag}}_k({x^{*}}_k) B(y)\Big \rangle \Big\langle B^{\dag}
(y^{*}){\cal O}_1(x_1){\cal O}_2(x_2)\cdots {\cal O}_k(x_k)\Big\rangle}{\big\langle B^{\dag}
(y^{*})B(y)\big\rangle}.
\label{overstatenorm1}
\eea
 Now it is straightforward to promote equation ({\ref{overstatenorm1}}) into the probability interpretation by
 dividing both sides of ({\ref{overstatenorm1}}) by the left hand side of the same equation and thus we get
\begin{eqnarray}
1 =
\sum_B \frac{\Big|\langle {{\cal O}^{\dag}}_1({x^{*}}_1){{\cal O}^{\dag}}_2
({x^{*}}_2)\cdots{{\cal O}^{\dag}}_k({x^{*}}_k) B(y) \rangle \Big|^2}
{ \big\langle {{\cal O}^{\dag}}_1({x^{*}}_1){{\cal O}^{\dag}}_2({x^{*}}_2)\cdots
{{\cal O}^{\dag}}_k({x^{*}}_k)
{\cal O}_1(x_1){\cal O}_2(x_2)\cdots {\cal O}_k(x_k)\big\rangle\big\langle B^{\dag}(y^{*})B(y)\big\rangle}.
\label{overstatnorm1}
\end{eqnarray}
We call $P$ as the probability for ${\cal O}_1(x_1){\cal O}_2(x_2)\cdots {\cal O}_k(x_k)$ to evolve into $B$.
\bea
&&P\Big({\cal O}_1(x_1){\cal O}_2(x_2)\cdots {\cal O}_k(x_k) \rightarrow B(y)\Big) =   \nonumber  \\
&&\frac{\Big| {{\cal O}^{\dag}}_1({x^{*}}_1){{\cal O}^{\dag}}_2
({x^{*}}_2)\cdots{{\cal O}^{\dag}}_k({x^{*}}_k) B(y) \rangle \Big|^2}
{\big \langle {{\cal O}^{\dag}}_1({x^{*}}_1){{\cal O}^{\dag}}_2({x^{*}}_2)\cdots
{{\cal O}^{\dag}}_k({x^{*}}_k)
{\cal O}_1(x_1){\cal O}_2(x_2)\cdots {\cal O}_k(x_k)\big\rangle\big\langle B^{\dag}(y^{*})B(y)\big\rangle}.
\label{overstatnorm1}
\eea
The above formula ({\ref{overstatnorm1}) for probability is based on the notion of overlap state normalization.
 If we replace the state $|B(y)\rangle$ by $|B(y_1)B(y_2)\rangle$, the probability will not correspond to the separate measurement
 of the operators $B(y_1)$ and $B(y_2)$.\\

\noindent
The extension of this idea for the probability interpretation of separate measurements or multi particle normalization needs correlator
 on higher topology. For two separate measurements, unlike the case of overlap state normalization we take a manifold $S^1\times S^2$ with
genus-1 and $2k$ number of operator insertions. Then cut out two boundaries $\mathbb{B}_1$ and $\mathbb{B}_2$ in such a way that both
manifolds have $k$ number of operator insertions as in fig \ref{torus}.
\begin{figure}[h]
\begin{center}
\begin{psfrags}
\psfrag{A}[][]{$\hspace{.3cm} =\sum_{B_1, B_2}$}
\psfrag{B}[][]{$\hspace{.1cm}\frac{|B_1\rangle\langle B_1|}{\langle B_1|B_1\rangle}$}
\psfrag{C}[][]{$\hspace{-.2cm}\frac{|B_2\rangle\langle B_2|}{\langle B_2|B_2\rangle}$}
\includegraphics[width=14cm]{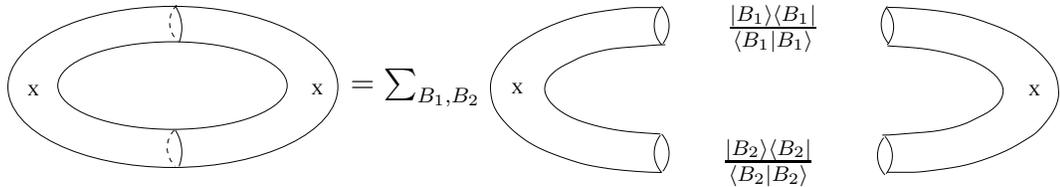}
\caption{In this figure k number of operator insertions represented by a single mark.}
\label{torus}
\end{psfrags}
\end{center}
\end{figure}

\noindent
Now for the n-point correlator, the factorization equation takes the following form
\begin{eqnarray}
&&\Big\langle{{\cal O}^{\dag}}_1({x^{*}}_1){{\cal O}^{\dag}}_2({x^{*}}_2)\cdots
{{\cal O}^{\dag}}_k({x^{*}}_k)
{\cal O}_1(x_1){\cal O}_2(x_2)\cdots {\cal O}_k(x_k)\Big\rangle _{G=1} = \nonumber  \\
&& \hspace{-.4in}\sum_{B_1,B_2} \frac{\Big\langle {{\cal O}^{\dag}}_1({x^{*}}_1)\cdots{{\cal O}^{\dag}}_k({x^{*}}_k)
 B_1(y_1)B_2(y_2)\Big\rangle \Big\langle {B^{\dag}}_2
(y_2^{*}){B^{\dag}}_1
(y_1^{*}){\cal O}_1(x_1)\cdots {\cal O}_k(x_k)\Big\rangle}{\big\langle {B^{\dag}}_1
(y_1^{*})B_1(y_1)\big\rangle \big\langle {B^{\dag}}_2
(y_2^{*})B_2(y_2)\big\rangle}
\label{multparttnorm1}
\end{eqnarray}
where $\cal O$'s are again the local operators defined on the genus-1, $S^1\times S^2$ manifold of interest whereas $B_1$'s and
$B_2$'s are the local operators defined on the boundary cut $\mathbb{B}_1$ and $\mathbb{B}_2$ respectively. Then the probability
interpretation arises in the same fashion
\begin{eqnarray}
&&P\Big({\cal O}_1(x_1){\cal O}_2(x_2)\cdots {\cal O}_k(x_k) \rightarrow B_1(y_1)B_2(y_2)\Big)
=\nn\\&& \hspace{-.4in}\sum_{B_1,B_2} \frac{\Big| {{\cal O}^{\dag}}_1({x^{*}}_1)\cdots{{\cal O}^{\dag}}_k({x^{*}}_k)
 B_1(y_1)B_2(y_2)\Big|^2 }{ \big\langle{{\cal O}^{\dag}}_1({x^{*}}_1)\cdot\cdot
{{\cal O}^{\dag}}_k({x^{*}}_k)
{\cal O}_1(x_1)\cdot\cdot {\cal O}_k(x_k)\big\rangle _{G=1}  \big\langle {B^{\dag}}_1
(y_1^{*})B_1(y)\big\rangle \big\langle {B^{\dag}}_2
(y_2^{*})B_2(y_2)\big\rangle}.
\label{multparttnorm2}
\end{eqnarray}
It is interesting to note that to formulate the probability interpretation for multiple number of separate measurements we need
higher genus factorization.\\

\noindent
Therefore depending on the number of measurements or number of out going states or operators we have to consider the appropriate
topology of the space on which the operators live to find out the transition probability among the states of the gravity. Keeping
this lesson in our mind we compute few probabilities in the forthcoming sections.

\section{Sphere Factorization}
In this section, we want to find out the probability of getting one state from the different number of states of the gravity.
Thus we should use the genus zero factorization. To do that first we consider two $S^3$ manifolds. By cutting out a 3-ball of unit
 radius around the origin of each one, we map them in two separate $\mathbb{R}^3$ spaces described by the set of coordinates
 $(r,\Omega_{2,r})$ and $(s,\Omega_{2,s})$ respectively. The metrics in the $\mathbb{R}^3$ spaces take the following forms,
\be
ds_r^2 = dr^2 + r^2d\Omega_{2,r}^2 \quad \quad {\rm and} \quad \quad ds_s^2 = ds^2 + s^2d\Omega_{2,s}^{ 2}.
\ee
Finally, we glue these two manifolds by using $rs=1$ to get the genus zero manifold of our interest. In this manifold, the most general formula for the probability can be written as
\bea
&&\hspace{-.5in}P\Big(R_1(r=e^{x_1})\cdots R_k(r=e^{x_k})\rightarrow R(r=0)\Big)\nn\\
&\hspace{-.7in}=&\hspace{-.4in}\frac{\big\langle R_1^\dagger(s=e^{x_1} ) \cdots R_k^\dagger(s=e^{x_k}) R(r=0) \big\rangle \big\langle  R^\dagger(s=0) R_k(r=e^{x_k})\cdots  R_1(r=e^{x_1})   \big\rangle}{\big\langle R_k^\dagger(s=e^{x_1}) \cdots R_1^\dagger(s=e^{x_k}) R_1(r=e^{x_k})\cdots  R_k(r=e^{x_1})\big\rangle \big\langle
    R^\dagger(s=0) R(r=0)\big\rangle}.
\label{spherefactin3d}
\eea
Notice that in this formula we have considered final operator at $r,s=0$ instead of $r=s=1$. To consider that we first replaced the operators at $r=s=1$ by their corresponding states via operator/state correspondence. We then evolved the states up to $r,s=0$ by doing the path integral over the unit disc. Since in these regions there are no operator insertions, in the end result states differ by a scale factor only. The scaling  factor of state in $r$ space is exactly canceled by the scale factor arises by evolving state in $s$ space. Finally we again replaced the states as their dual operators at $r,s=0$ (for details see \cite{Brown}). \\

\noindent
Before going to find out the probability for a specific case, we first compute few useful correlators which will be used in
our later computations. From now we suppress the angular dependence part in most of the computations since the angles, in all of
 the gluings, are identified trivially. Most of the time, we also abbreviate $\chi_R(AB^\dagger)$ and $\chi_R(A^\dagger B)$ as $R$
and $R^\dagger$. First we like to find out the two point function of two Schur polynomials, one in each $S^3$ with a cut-out of 3-ball
those are glued together to construct the manifold of our interest. The two point function of interest is
\begin{equation}
\langle R^{\dagger}(s=0)R(r=0)\rangle
\end{equation}
with $s=1/r$ and $R^{\dagger}$ is a conformal primary operator. To calculate the correlator we should bring the operator
 $R^{\dagger}$ in the $r$ coordinate frame. Under this coordinate change, the metric of $\mathbb{R}_s^{3}$ changes as
\begin{equation}
ds^{ 2} + s^{2}d\Omega_{2,s}^{2} \rightarrow \frac{1}{r^4}(dr^{2} + r^{2}d\Omega_{2,r}^2)
\end{equation}
and so the operator $R^{\dagger}$ transform as
\begin{equation}
R^{\dagger}(s)\rightarrow \Omega(r)^{-\Delta/2}R^\dagger(r)= r^{2\Delta}R^\dagger(r)
\end{equation}
where $\Omega(r)=1/r^4$ and $\Delta$ is the conformal dimension of the operator $R^\dagger$. Now we are ready to compute the
above correlator which is also called Zamalodchikov metric of the conformal field theory. The  result is
\begin{eqnarray}
\nonumber\langle R^{\dagger}(s=0)R(r=0)\rangle &=& \lim_{r_0\rightarrow\infty}\langle r_0^{2\Delta} R^{\dagger}(r=r_0)R(r=0)\rangle\\\nonumber
&=&\lim_{r_0\rightarrow\infty}\Big[ \frac{r_0^{2\Delta}f_R^{N_1}f_R^{N_2}}{(4\pi|0 -r_0|)^{2\Delta}}\Big]\\
&=&\frac{f_R^{N_1}f_R^{N_2}}{(4\pi)^{2\Delta}}.
\end{eqnarray}
We would also like to find out this correlator:
\be
A_M^1=\langle\Tr(AB^\dagger)^M\Tr(A^\dagger B)^M\rangle.
\ee
We find out this by writing down a recursion relation. Choose one trace from $\Tr(AB^\dagger)^M$ and contract the involved fields $A$
and $B^\dagger$ with the fields $A^\dagger$ and $B$ of the other trace $\Tr(A^\dagger B)^M$. We can do this in two ways: firstly, $A$
and $B^\dagger$ contract with $A^\dagger$ and $B$ fields of the same trace of $\Tr(A^\dagger B)^M$ and give $N_1 N_2$. There lies $M$
number of choices since there is $M$ number of traces. In the second way, $B^\dagger$ does not contract with $B$ field of trace
where $A^\dagger$ is already contracted with $A$. So for this way, $A$ field has again $M$ number of choices but for the $B^\dagger$
there are $(M-1)$ choices. For both ways, at the end $(M-1)$ number of traces will be left out. So we can write the above correlator
through this recursion relation
\bea
A_M^1 &=& \Big(MN_1 N_2 + M(M-1)\Big) A_{M-1}\nn\\
&=&M(N_1 N_2 +M -1)A_{M-1}\nn\\
&=&M!\frac{(N_1 N_2 +M -1)!}{(N_1 N_2 -1)!}.
\eea
Finally, before going to the main computation, we compute the correlator like
\be
A_M^J=\langle\Tr((AB^\dagger)^J)^M\Tr((A^\dagger B)^J)^M\rangle.
\ee
Since exact computation is difficult, we restrict the analysis up to leading order where $JM << N_1 N_2$ and the
result is
\be
A_M^J=\langle\Tr((AB^\dagger)^J)^M\Tr((A^\dagger B)^J)^M\rangle = J^M M! (N_1N_2)^{JM}.
\ee
By using the above correlators, we first calculate the probability to get one giant graviton from
two giant gravitons of angular momentum $N_1/2$. The formula in
equation(\ref{spherefactin3d}) reduces to
\bea
\nonumber P\hspace{-.4in}&&\Big(R_1(r=e^x),R_2(r=e^x)\rightarrow R(r=0)\Big)\\\nonumber
&=&\frac{{\Big|}\langle R_1^\dagger(r=e^x)R_2^\dagger(r=e^x) R(r=0)\rangle{\Big|}^2}{\langle R_2^\dagger
(s=e^x)R_1^\dagger(s=e^x)R_1(r=e^x)R_2(r=e^x)\rangle\langle R^\dagger(s=0) R(r=0)\rangle}\\
& = & \frac{g\left(R_1,R_2;R\right)^2\Big[f_{R}^{N_1}f_{R}^{N_2}\Big]^2e^{-4N_1x} (4\pi)^{-4N_1}}{\sum_S g\left(R_1,R_2;S\right)^2 f_S^{N_1}f_S^{N_2}e^{-2N_1x}(4\pi)^{-2N_1}(e^x - e^{-x})^{-2N_1} f_{R}^{N_1}f_{R}^{N_2}(4\pi)^{-2N_1} }.
\eea
Here we have considered $R_1$ and $R_2$ at the same position, so that the normalization factor in the denominator is an
extremal correlator. We also consider the large $x$ limit to maximize the distance between the operators $R_1$ and $R_2$
from $R$. It gives the probability which is space-time independent. Thus, at large $x$ limit, we get
\be
  P(R_1,R_2 \to R) = \frac{f_{R}^{N_1}f_{R}^{N_2}}{\sum_S g\left(R_1,R_2;S\right)^2 f_S^{N_1} f_S^{N_2}  }.
\ee
The fusion of the two sphere giants (two vertical Young diagrams of length $N_1/2$) gives a sum of
representations, with column lengths   $  ( N_1/2 + i , N_1/2 -i )$. Hence the denominator can be written as
\be
\sum_{ i,j = 0}^{N_1/2 }  { N_1! ( N_1+1)! \over  ( N_1/2 -i )! ( N_1/2 + i + 1 ) !}{N_2! ( N_2 +1)!\over( N_2-N_1/2 -j )! (N_2 - N_1/2 + j + 1 ) !}.
\ee
By straightforward simplification we arrives at the final expression for the probability of sphere giant
\be
\sim e^{-\big[(2N_2+\frac{3}{2})\log(1+\frac{1}{N_2}) +
(N_2-N_1+\frac{1}{2})\log(1-\frac{N_1}{N_2})+ (N_1-2N_2-\frac{3}{2})\log(1-\frac{N_1}{2N_2})+ 2N_1\log(2)-2\big]}.
\ee
Similarly for AdS giant, the denominator will be
 \be
\sum_{ i,j = 0}^{N_1/2 } {  ( 3N_1/2  + i -1 ) ! ( 3N_1/2 -i -2 ) ! \over ( N_1-1) ! ( N_1-2) ! }{  ( N_2 +N_1/2  + j -1 ) !
( N_2+N_1/2 -j -2 ) ! \over ( N_2-1) ! ( N_2-2) ! }.
\ee
Since, in the sum the $f_{R}^{N_1}f_{R}^{N_2}$ is included, the probability becomes less than 1.\\

\noindent
 Let us consider one sphere giant graviton produced by the combination of $N_1$ number of KK gravitons of angular momentum 1.
From now, we assume that $N_1$ is less than $N_2$. Since these KK gravitons are with angular momentum 1,
we should write them in terms of trace operators. Thus we write the probability as
\bea
\nonumber P\hspace{-.4in}&&\Big(\Tr(AB^\dag)^{N_1}(r=e^x)\rightarrow \chi_{[1^{N_1}]}(AB^\dag)(r=0)\Big)\\
&=&\frac{{\Big|}\langle \Tr(A^\dag B)^{N_1}(r=e^x)\chi_{[1^{N_1}]}(A B^\dag) (r=0)\rangle{\Big|}^2}{\langle
\Tr(A^\dag B )^{N_1}(s=e^x)\Tr(A B^\dag)^{N_1}(r=e^x)\rangle\langle \chi_{[1^{N_1}]}(A^\dag B)
\chi_{[1^{N_1}]}(A B^\dag)\rangle}.
\eea
Since in the  numerator one side of the correlator is in Schur basis, we should change our trace operator in to Schur polynomial
 and we do that by the formula derived in\cite{SS}
\be
\Tr((A^\dagger B)^n) = \sum_R \chi_R(\sigma)\chi_R(A^\dagger B)
\label{trtoschur}
\ee
where $R$ is the representation of the symmetric group $S_n$ and $\chi_R(\sigma)$ is the character of a cycle of length $n$\cite{Fulton}.
  By doing this one can write the trace operator with unit $\cal R$-charge that corresponds to a single box Young diagram as
$$\Tr(A^\dag B)= \chi_R(A^\dag B).$$
Therefore we can write $N_1$ number of KK graviton with angular momentum 1 as
\be
\Big(\Tr(A^\dagger B)\Big)^{N_1}= \Big(\chi_R(A^\dagger B)\Big)^{N_1}= \sum_S g(R_1\cdots R_{N_1}; S)\chi_S(A^\dagger B).
\ee
Finally, by replacing trace basis as Schur basis in the numerator, we have the probability
\bea
\nonumber P\hspace{-.4in}&&\Big(\Tr(A B^\dag)^{N_1}(r=e^x)\rightarrow \chi_{[1^{N_1}]}(A B^\dag)(r=0)\Big)\\\nonumber
&=&\frac{{\Big|}\langle \sum_{S} g(R_1\cdots R_{N_1};S)\chi_{S}(A^\dag B)(r=e^x)\chi_{[1^{N_1}]}(A B^\dag) (r=0)\rangle{\Big|}^2}{\langle
\Tr(A^\dagger B)^{N_1}(s=e^x)\Tr(AB^\dagger)^{N_1}(r=e^x)\rangle\langle \chi_{[1^{N_1}]}(A^\dagger B) \chi_{[1^{N_1}]}(AB^\dagger )\rangle}\\\nn
&=& \frac{\Big[f_{[1^{N_1}]}^{N_1}f_{[1^{N_1}]}^{N_2}\Big]^2e^{-4N_1x} (4\pi)^{-4N_1}}{N_1!\frac{(N_1N_2 +N_1-1)!}{(N_1N_2-1)!}e^{-2N_1x}(4\pi)^{-2N_1}(e^x - e^{-x})^{-2N_1}f_{[1^{N_1}]}^{N_1}f_{[1^{N_1}]}^{N_2}(4\pi)^{-2N_1} }.
\eea
In the second line we use the fact that two point function is finite when representation $S$ is same with the representation of
 the sphere giant. Thus the contributing Littlewood-Richardson number is only $g(R_1\cdots R_{N_1};{[1^{N_1}]})$ which has value 1.
 Again at large $x$ limit probability reduces to
\bea
\nonumber P\hspace{-.4in}&&\Big(\Tr(AB^\dagger )^{N_1}(r=e^x)\rightarrow \chi_{[1^{N_1}]}(AB^\dagger)(r=0)\Big)\\
&=&\frac{\Big[f_{[1^{N_1}]}^{N_1}f_{[1^{N_1}]}^{N_2}\Big]}{N_1!\frac{(N_1N_2 +N_1-1)!}{(N_1N_2-1)!} }=\frac{\frac{N_1!N_2!}{(N_2-N_1)!}}{N_1!\frac{(N_1N_2+N_1-1)!}{(N_1N_2-1)!}}=\frac{N_2!}{(N_2-N_1)!}\frac{(N_1N_2-1)!}{(N_1N_2 +N_1-1)!}\nn\\
&\sim & e^{-[N_1 + \frac{N_1}{N_2}-\frac{1}{2N_2} +N_1\log (N_1) + (N_2-N_1 +\frac{1}{2})\log(1 - \frac{N_1}{N_2})]}.
\eea
 Similarly to produce an AdS giant from $N_1$ number of KK gravitons with angular momentum 1 we will have the probability
\bea
\nonumber P\hspace{-.4in}&&\Big(\Tr(AB^\dagger)^{N_1}(r=e^x)\rightarrow \chi_{[N_1]}(AB^\dagger)(r=0)\Big)\\
&=&\frac{\frac{(2N_1-1)!(N_2+N_1-1)!}{(N_1-1)!(N_2-1)!}}{N_1!\frac{(N_1N_2 +N_1-1)!}{(N_1N_2-1)!}}=\frac{(2N_1-1)!(N_2+N_1-1)!(N_1N_2-1)!}{(N_1-1)!(N_2-1)!N_1!(N_1N_2 +N_1-1)!}\nn\\
&\sim& \pi^{-\frac{1}{2}}e^{-\big[N_1 + \frac{N_1}{N_2}-\frac{1}{2N_2}+(N_1+\frac{1}{2})\log(N_1)  - (N_1 + N_2-\frac{1}{2})\log(1+ \frac{N_1}{N_2})-(2N_1-1)\log(2)\big]}.
\eea
 \noindent
 Now consider a combination of $L/J$ number of gravitons with angular momentum $J<\sqrt N_1$, resulting into a sphere giant of angular
 momentum $L$.
 So the probability for this transition will be
 \bea
\nonumber & \hspace{- 2 cm} P\Big(\Tr((AB^\dagger)^J)^{L/J}(r=e^x)\rightarrow \chi_{[1^{L}]}(AB^\dagger)(r=0)\Big)\\\nonumber
=& \frac{{\Big|}\langle \Tr((A^\dagger B)^J)^{L/J}(r=e^x)\chi_{[1^{L}]}(AB^\dagger) (r=0)\rangle{\Big|}^2}
{\langle \Tr((A^\dagger B)^J)^{L/J}(s=e^x)\Tr((AB^\dagger)^J)^{L/J}(r=e^x)\rangle\langle \chi_{[1^{L}]}(A^\dagger B)
\chi_{[1^{L}]}(AB^\dagger)\rangle}\\
=& \frac{\sum_{R_1.. R_{L/J}} g(R_1,.., R_{L/J};[1^{L}])^2\big[\chi_{R_1}\left(J\right)..
 \chi_{R_{L/J}}\left(J\right)\big]^2\Big[f_{[1^{L}]}^{N_1}f_{[1^{L}]}^{N_2}\Big]^2e^{-4Lx} (4\pi)^{-4L}}{ J^{L/J}(L/J)!
(N_1N_2)^{L} e^{-2Lx}(4\pi)^{-2L}(e^x - e^{-x})^{-2L} f_{[1^{L}]}^{N_1}f_{[1^{L}]}^{N_2}(4\pi)^{-2L}}.
 \eea
 In this calculation we consider the leading order value of the first correlator of the denominator.
 Since the final state is sphere giant, which is an antisymmetric representation, the only allowed representations of
the gravitons are antisymmetric so that combined representation gives the antisymmetric representation. Therefore,
we do not need the sum over Littlewood-Richardson number and it gives 1 only and also $\chi_R(J)$, the characters of a
cycle of length $J$, always be $\pm 1$ and due to square over the product of the characters the total contribution of them
will be 1 only. Thus at the large $x$ limit, the probability reduces to
 \bea
 & \frac{\Big[f_{[1^{L}]}^{N_1}f_{[1^{L}]}^{N_2}\Big]}{ J^{L/J}(L/J)!(N_1N_2)^{L} }
=\frac{N_1!N_2!}{ J^{L/J}(L/J)!(N_1N_2)^{L} (N_1-L)!(N_2-L)!}\nn\\
 \sim & \pi^{-\frac{1}{2}} e^{-[2L-\frac{L}{J}-\frac{1}{2}\log (J)+
 (\frac{L}{J}+\frac{1}{2})\log(L)+(N_1 - L+\frac{1}{2})\log(1-\frac{L}{N_1}) +(N_2 - L+\frac{1}{2})\log(1-\frac{L}{N_2})+
 \frac{1}{2}\log(2)]}.
\eea
Similarly, to create an AdS giant from $L/J$ number of gravitons with angular momentum $J<\sqrt N_1$, the probability will be
 \bea
\nonumber & P\Big(\Tr((AB^\dagger)^J)^{L/J}(r=e^x)\rightarrow \chi_{[{L}]}(AB^\dagger)(r=0)\Big)
= \frac{(N_1+L-1)!(N_2+L-1)!}{ J^{L/J}(L/J)!(N_1N_2)^{L} (N_1-1)!(N_2-1)!}\\
\sim & \pi^{-\frac{1}{2}} e^{-[2L-\frac{L}{J}-\frac{1}{2}\log (J)+ (\frac{L}{J}+\frac{1}{2})
\log(L)-(N_1 + L-\frac{1}{2})\log(1+\frac{L}{N_1}) -(N_2 + L-\frac{1}{2})\log(1+\frac{L}{N_2})+ \frac{1}{2}\log(2)]}.
\eea
All these correlators are always less than 1 and decay exponentially with $N_1$ and $N_2$.\\

\noindent
If $L$ number of gravitons with $J$ amount of angular momentum interact and produce single graviton of angular momentum $(LJ)$
which is less than $\sqrt N_1$, the probability can be written as
\bea
\nonumber & P\Big(\Tr((AB^\dagger)^J)^{L}(r=e^x)\rightarrow \Tr((AB^\dagger)^{LJ})(r=0)\Big) \\\nonumber
=&\frac{{\Big|}\langle \Tr((A^\dagger B)^J)^{L}(r=e^x)\Tr((AB^\dagger)^{LJ}) (r=0)\rangle{\Big|}^2}{\langle \Tr((A^\dagger B)^J)^{L}
(s=e^x)\Tr((AB^\dagger)^J)^{L}(r=e^x)\rangle\langle  \Tr((A^\dagger B)^{LJ})\Tr((AB^\dagger)^{LJ})\rangle}\\
=&\frac{\sum_{R_1 \cdots R_L;S} g(R_1, \dots, R_L;S)^2\big[\chi_{R_1}\left(J\right) \cdots
 \chi_{R_L}\left(J\right)\chi_{S}\left(LJ\right)\big]^2\Big[f_S^{N_1}f_S^{N_2}\Big]^2e^{-4LJx} (4\pi)^{-4LJ}}{ J^{L}L!(N_1N_2)^{LJ} e^{-2LJx}(4\pi)^{-2LJ}(e^x - e^{-x})^{-2LJ} (LJ)!\frac{(N_1N_2 + LJ -1)!}{(N_1N_2 -1)!}(4\pi)^{-2LJ}}.\\\nn
\eea
Since we know $\chi_R(I)$ will only be non-zero for hooks $\chi_{[(R-r),1^r]}(I)=(-1)^r$.
Therefore the contribution of the each character is $\pm 1$.  Thus the total contribution of the characters is only 1.
 Then the above probability reduces at large $x$ limit as
\bea
\nonumber & P \Big(\Tr((AB^\dagger)^J)^{L}(r=e^x)\rightarrow \Tr((AB^\dagger)^{LJ})(r=0)\Big)\\
=&\frac{\sum_{r_1\cdots r_L,s} g\Big([(R_1-r_1),1^{r_1}], \dots,[(R_L-r_L),1^{r_L}] ;[(S-s),1^{s}]\Big)^2
\Big[f_{[(S-s),1^{s}]}^{N_1}f_{[(S-s),1^{s}]}^{N_2}\Big]^2}{ J^{L}L!(N_1N_2)^{LJ}(LJ)!\frac{(N_1N_2 + LJ -1)!}{(N_1N_2 -1)!}}.
\eea
Using this method one can find out other type of correlators those will produce only one final state. Like large number of gravitons with different $\cal R$-charge producing one sphere giant, AdS giant or graviton. However we are not going to find out those in this paper. We close this section with these above correlators. In the next section we compute the correlators for two out going states.

\section{The genus one factorization}

\noindent
To obtain the probability of finding two states from many states we need to compute correlators on genus one manifold.
Particularly, we consider here the manifold $S^1\times S^2$.
We construct this manifold by gluing two cylinders $I\times S^2$ described by coordinates $(r,\Omega_{2,r})$ and $(s,\Omega_{2,s})$
 in the two different $\mathbb{R}^3$ spaces. The radial variables are bounded by these ranges
\be
1\le r\le e^T \quad\quad {\rm and}\quad\quad 1\le s\le e^T.
\ee
We also introduce the coordinates $r^\prime=1/r$ and $s^\prime=1/s$. Now to produce the manifold $S^1\times S^2$, we glue the two cylinders at the inner ends $r=1$, $s=1$ with $rs=1$ and outer ends at $r=e^T$, $s=e^T$ with $r^\prime s^\prime=e^{-2T}({\rm i.e}\; rs =e^{2T} )$. \\

\noindent
As earlier, we can now define the probability of one giant graviton going to two smaller giant gravitons as follows
\bea
 && P\Big(R(r=e^x) \to R_1'(r'=0) R_2(r=0)\Big)\nn \\
 & =& \frac{\left|\big\langle
       R^\dagger(r=e^x) R_1'(r'=0)R_2(r=0) \big\rangle
    \right|^2}{\big\langle
    R^\dagger (s=e^x) R(r=e^x)\big\rangle_{G=1}  \big\langle
    {R_1^{\dagger}}^{\prime}(s^\prime=0)R_1^\prime(r^\prime=0)\big\rangle\big\langle R_2^\dagger(s=0) R_2(r=0)\big\rangle}.
    \label{probability1}
\eea
Following the same logic as described in the previous section we have again considered the operators at $r, r^\prime =0$.
To compute this probability we study term by term. Lets first work out the three point function of the numerator
\bea
& &\langle R^{\dagger} ( r = e^x ) R_1^\prime ( r'  = 0 )
   R_2  ( r =0 ) \rangle \nn \\
& =& \lim_{r_0 \to \infty} \langle R^{\dagger  } ( r = e^x ) r_0^{2\Delta_1} R_1 ( r = r_0 )
   R_2  ( r =0 ) \rangle \nn \\
& = &(4\pi)^{-2(\Delta_1 +\Delta_2)} e^{-2x\Delta_2}g(R_1,R_2;R) f_R^{N_1}f_R^{N_2}.
\eea
Then we compute second two point function of the denominator
\be
\langle {R_1^{\dagger}}^{\prime}(s^\prime=0)R_1^\prime(r^\prime=0)\rangle = (4\pi)^{-2\Delta_1}e^{2T\Delta_1}f_{R_1}^{N_1}f_{R_1}^{N_2}.
\ee
Similarly for the third two point function of the denominator we get
\be
\langle R_2^\dagger(s=0) R_2(r=0)\rangle = f_{R_2}^{N_1}f_{R_2}^{N_2}(4\pi)^{-2\Delta_2}.
\ee
 In addition to these three correlators we need to know one more correlator which is on $S^2 \times S^1$, seating in the denominator .
 For the space-time dependent part of the correlator, we have to know the Green's function for this manifold.
 To find out Green's function we start with the associated metric of this space
\be
ds^2 = d{\tau}^2 + d{\theta}^2 + \sin^2\theta d\phi^2
\ee
with the proper range of coordinates, $\tau \in [0, 2T]$, $\theta \in [0,\pi]$ and $\phi \in [0,2\pi]$.\\

\noindent
Sturm-Liouville theory suggests that, if the eigenvectors $\Psi_n(x)$ of a hermitian operator $\mathcal{L}$ span a  basis,
the Green's function of interest, $G(x,y)$ is expressible as a linear combination of the $\Psi_n(x)$
\be
G(x,y) = \sum_{n|\lambda_n\neq 0} \frac{{\Psi^{\ast}_n}(x)\Psi_n(x)}{\lambda_n}.
\ee
As we are analysing three dimensional conformal filed theory of scalar field on 3-sphere , while defining the differential
operator $\mathcal{L}$ it is necessary to consider coupling to the 3-dimensional curvature. In general  $\mathcal{L}$  takes the
 following form
\be
\mathcal{L}= \Delta - \frac{R}{8}
\ee
where $\Delta$ is Laplacian and not conformal dimension. $R$ is the Ricci scalar. More specifically,
considering the space $S^1\times S^2$ with unit radii and also noting that only the curvature of $S^2$ contributes,
the Ricci scalar comes out as $R = 2$. The differential operators $\mathcal{L}$ modifies into a particular form
\be
\mathcal{L} = \Delta_{Euclidean} - \frac{1}{4}.
\ee
The form of differential operator $\mathcal{L}$ leads to the identification of its eigenvectors as a complete set of spherical
harmonics on $S^2 \times S^1$
\be
\Psi_n = \varsigma_m(\tau)Y^M_J(\theta,\phi)
\ee
where $n = (m, J, M)$. The explicit form of $S^2$ harmonics is given by
\be
Y^M_J(\theta,\phi) = \sqrt{\frac{(2J+1)}{4\pi}}\frac{(J-M)!}{(J+M)!} (-1)^{M}\sin^M \theta {\Big(\frac{d^{J+M}}{{d^{J+M}\cos\theta}}\Big)}
\frac{{(\cos^2\theta - 1)}^J}{2^J J!}
\label{sphericalh}
\ee
where the quantum numbers $J$ and $M$ take values as
\bea
&& \hspace{-1.5in} J = 0,1,2,3\cdots \nn\\
M = -J,-J+1,\cdots J.
\eea
The harmonics on $S^1$ are
\be
\varsigma_m(\tau) = \frac{1}{\sqrt{2T}} e^{\frac{i m \pi \tau}{T}},
\ee
with $m$ takes the values as, $m = 0,1,2\cdots$.\\
The spherical harmonics of $S^2$ and $S^1$ satisfy the following differential equations respectively
\bea
&& \hspace{-2 in} \Delta_{S^2} Y^M_J(\theta,\phi) = -J(J+M)Y^M_J(\theta,\phi), \nn \\
 \Delta_{S^1}  = -{\Big(\frac{m\pi}{T}\Big)}^2 \varsigma_m(\tau).
\eea
Now with all the eigenfunctions of Euclidean Laplacian operators defined on $S^2\times S^1$ in hand, we are able to write the differential equation for operator $\mathcal{L}$
\be
\mathcal{L} \Psi_n  = \big(\Delta_{S^2 \times S^1} - \frac{1}{4}\big) \Psi_n = \Big[-J(J+M)-{\Big(\frac{m\pi}{T}\Big)}^2-\frac{1}{4}\Big]\Psi_n.
\ee
In the 3-dimensional space-time of our interest we define the action of operator $\mathcal{L}$ on corresponding Green's function defined in
conformity with $\mathbb{R}^3$ correlator.
\be
 \mathcal{L} G(x,y) = -\delta^3(x-y)
\ee
and the Green's function in terms of the spherical harmonics is
\be
  G(x,y) = \sum_{J,M,m} \frac{{\varsigma_m}^\ast(\tau) {Y^M_J}^\ast(\theta,\phi){\varsigma_m}(\tau) {Y^M_J}(\theta,\phi)}{J(J+M) +
 {(\frac{m\pi}{T})}^2 + \frac{1}{4} }.
\ee
Now we are ready to compute the two point correlator on $S^1\times S^2$
\be
\langle R^{\dagger}(s = e^x) R(r = e^x)\rangle _{G=1}.
\ee
Since the associated metric of  $S^1\times S^2$ involves the coordinate $\tau, \theta, \phi$ and we suppress angular coordinates in our
calculation, we should bring the correlator in $\tau$ coordinate instead of $r\; {\rm and}\; s$. We do that by changing the coordinates
$s = e^{-\tau},\; r = e^\tau$ and finally we get
\be
\langle R^{\dagger}(s = e^x) R(r = e^x)\rangle _{G=1} = e^{-2x \Delta} {\langle R^{\dagger}(\tau = -x) R(\tau = x)\rangle}_{G = 1}
\ee
with $\Delta = \Delta_1 + \Delta_2$ by charge conservation. Finally in terms of gauge index and Green's function we have the correlator as
\be
\langle R^{\dagger}(s = e^x) R(r = e^x)\rangle _{G=1} = e^{-2x \Delta} f_R^{N_1}f_R^{N_2}\Big[\sum_{J,M,m}
\frac{{\varsigma_m}^\ast(0) {Y^M_J}^\ast(\theta,\phi){\varsigma_m}(2x) {Y^M_J}(\theta,\phi)}{J(J+M) +
 {(\frac{m\pi}{T})}^2 + \frac{1}{4} }\Big]^{2\Delta}.
\ee
As we suppress angular part, the final result should be independent of the choice of $\theta$ and $\phi$. We thus choose $\theta = 0$, so that the sum simplifies significantly. Equation(\ref{sphericalh}) demands for $\theta = 0$ the only non-zero term contributing corresponds to $J=0,M=0$ and the above correlator reduces to
\be
\langle R^{\dagger}(s = e^x) R(r = e^x)\rangle _{G=1} = e^{-2x \Delta} f_R^{N_1}f_R^{N_2}\Big[\frac{1}{2T}\sum_{m} e^{\frac{im2\pi x}{T}}
\frac{ {Y^0_0}^\ast(\theta,\phi) {Y^0_0}(\theta,\phi)}{
 {(\frac{m\pi}{T})}^2 + \frac{1}{4} }\Big]^{2\Delta}.
\ee
With the value of harmonic function $Y^0_0 = 2^{-1}\pi^{-\frac{1}{2}} $ we get the final expression
\bea
\langle R^{\dagger}(s = e^x) R(r = e^x)\rangle _{G=1}&=& \Big[\frac{1}{8\pi T}[2\sum_{m > 0}
\frac{\cos \frac{m2\pi x}{T}}{(\frac{m\pi}{T})^2 +\frac{1}{4}} + 4]\Big]^{2\Delta} e^{-2x \Delta} f_R^{N_1}f_R^{N_2}\nn\\
&=& \Big[\frac{1}{8 \pi T}[8 \sum_{m > 0}
\frac{(-1)^m}{(\frac{2 m\pi}{T})^2 + 1} + 4]\Big]^{2\Delta} e^{-2x \Delta} f_R^{N_1}f_R^{N_2}\nn\\
&=& \Big[\frac{1}{8 \pi T} [-4 + T(\coth \frac{T}{4} - \tanh \frac{T}{4} ) + 4 ]\Big]^{2\Delta} e^{-2x \Delta} f_R^{N_1}f_R^{N_2}\nn\\
&=& \Big[\frac{1}{4\pi} {\text{cosech}}\frac{T}{2}\Big]^{2\Delta} e^{-2x \Delta} f_R^{N_1}f_R^{N_2}\nn\\
&=& \Big[\frac{1}{4\pi} (2e^{-T/2})\Big]^{2\Delta} e^{-2x \Delta} f_R^{N_1}f_R^{N_2}.
\eea
To get the last line we have used the large $T$ limit where ${\text{cosech}}\frac{T}{2}\rightarrow 2e^{-T/2}$.\\

\noindent
Thus combining all four separate correlators we get the probability of one giant graviton goes to two giant gravitons as
\bea
 && P\Big(R(r=e^{T/2}) \to R_1(r=e^T) R_2(r=0)\Big)\nn \\
 &=&\frac{g(R_1,R_2;R)^2 f_R^{N_1}f_R^{N_2}}{2^{2(\Delta_1 + \Delta_2)}f_{R_1}^{N_1}f_{R_1}^{N_2}f_{R_2}^{N_1}f_{R_2}^{N_2}}
 \eea
in the  $T\rightarrow \infty$ limit and $R$ at $r=e^{T/2}$, which maximize the distance of the operators $R_1$ and $R_2$ from $R$ and suppress the space-time dependence in the probability as earlier.\\

\noindent
In particular, the probability of the transition of an AdS giant with angular momentum $N_1$ into to two smaller AdS giants
with angular momentum $N_1/2$  is given by
\bea
&& \frac{1}{2^{2N_1}} \frac{f_{[N_1]}^{N_1}f_{[N_1]}^{N_2}}{\Big[f_{[N_1/2]}^{N_1}f_{[N_1/2]}^{N_2}\Big]^2}
  = \frac{1}{2^{2N_1}} \frac{(2N_1-1)!(N_1-1)!(N_2+N_1-1)!(N_2-1)!}{\left[(3N_1/2-1)!(N_2 +N_1/2-1)!\right]^2} \nn\\
&\sim& e^{-\big[(2N_2 +N_1 -1)\log(1+ \frac{N_1}{2N_2})- (N_2 +N_1 -\frac{1}{2})\log(1+ \frac{N_1}{N_2})+(3N_1 - 1)\log(3)- (3N_1 - \frac{3}{2})\log(2)\big]}.
\eea
For a sphere giant $[1^N]$ evolving into two smaller sphere giants $[1^{\frac{N}{2}}]$ this probability becomes
\bea
&&\frac{1}{2^{2N_1}} \frac{f_{[1^{N_1}]}^{N_1}f_{[1^{N_1}]}^{N_2}}{\Big[f_{[1^{N_1/2}]}^{N_1}f_{[1^{N_1/2}]}^{N_2}\Big]^2}
 = \frac{1}{2^{2N_1}} \frac{\Big[(N_1/2)!(N_2-N_1/2)!\Big]^2}{N_1!N_2!(N_2-N_1)!} \nn\\
 &\sim& \pi^{\frac{1}{2}}e^{-\big[(N_2 -N_1 +\frac{1}{2})\log(1- \frac{N_1}{N_2})- (2N_2 -N_1 +1)\log(1- \frac{N_1}{2N_2})-\frac{1}{2}\log(N_1)+(3N_1 + \frac{1}{2})\log(2)\big]}.
\eea
We can also compute the transition probability of one sphere giant with angular momentum $N_1$ going into two gravitons with angular momentum $N_1/2$. Where $N_1/2$ is less than $\sqrt N_1$. In this case we can consider the trace basis for gravitons as genus zero case and then transform as Schur basis by using the equation(\ref{trtoschur}). Then the probability reduces to
\bea
 && P\Big(\chi_{[1^{N_1}]}(AB^\dag)(r=e^{T/2}) \to \Tr((AB^\dagger)^{N_1/2})(r=e^T) \Tr((AB^\dagger)^{N_1/2})(r=0)\Big)\nn \\
&=&\frac{\sum_{R_1, R_2} g(R_1, R_2;[1^{N_1}])^2\big[\chi_{R_1}\left(N_1/2\right)
 \chi_{R_2}\left(N_1/2\right)\big]^2\Big[f_{[1^{N_1}]}^{N_1}f_{[1^{N_1}]}^{N_2}\Big]}{ 2^{2N_1}(N_1/2)^2(N_1N_2)^{N_1} }\nn\\
 &=&\frac{N_1!N_2!}{2^{2N_1}(N_1/2)^2(N_1N_2)^{N_1}(N_2-N_1)!}\nn\\
 &\sim& \pi^{\frac{1}{2}}e^{-\big[2N_1 + \frac{3}{2}\log(N_1)+(N_2 -N_1 +\frac{1}{2})\log(1- \frac{N_1}{N_2})+(2N_1-\frac{5}{2})\log(2)\big]}.
 \eea
To get the 3rd line from 2nd line we have again used the fact that we can only have the antisymmetric representations of the
 gravitons from a sphere giant and in this condition sum over Littlewood-Richardson number takes the value 1. Same computation can
be done for the AdS giant also. In that case probability becomes
 \bea
 && P\Big(\chi_{[N_1]}(AB^\dag)(r=e^{T/2}) \to \Tr((AB^\dagger)^{N_1/2})(r=e^T) \Tr((AB^\dagger)^{N_1/2})(r=0)\Big)\nn \\
&=&\frac{\Big[f_{[{N_1}]}^{N_1}f_{[{N_1}]}^{N_2}\Big]}{ 2^{2N_1}(N_1/2)^2(N_1N_2)^{N_1} }\nn\\
 &=&\frac{(2N_1-1)!(N_2+N_1-1)!}{2^{2N_1}(N_1/2)^2(N_1N_2)^{N_1}(N_2-1)!(N_1-1)!}\nn\\
 &\sim & e^{-\big[2N_1 +2\log(N_1)-(N_2 +N_1 -\frac{1}{2})\log(1+ \frac{N_1}{N_2})-\frac{3}{2}\log(2)\big]}.
 \eea
Again all correlatots are less than 1 and decaying exponentially with $N_1$ and $N_2$.\\

\noindent
If $L$ number of gravitons with $J$ amount of angular momentum interact and produce two gravitons of angular momentum $L_1$ and $L_2$
which are less than $\sqrt N_1$, the probability takes the following form
\bea
\nonumber P\hspace{-.4in}&&\Big(\Tr((AB^\dagger)^J)^{L}(r=e^x)\rightarrow \Tr((AB^\dagger)^{L_1})(r=e^T)\Tr((AB^\dagger)^{L_2})(r=0)\Big)\\\nonumber
&=&\frac{{\Big|}\langle \Tr((A^\dagger B)^J)^{L}(r=e^x)\Tr'((AB^\dagger)^{L_1}) (r'=0)\Tr((AB^\dagger)^{L_2}) (r=0)\rangle{\Big|}^2}{\langle \Tr((A^\dagger B)^J)^{L}(s=e^x)\Tr((AB^\dagger)^J)^{L}(r=e^x)\rangle_{G=1}}\\\nn &\times& \frac{1}{\langle  \Tr'((A^\dagger B)^{L_1})(s'=0)\Tr'((AB^\dagger)^{L_1})(r'=0)\rangle \langle  \Tr((A^\dagger B)^{L_2})(s=0)\Tr((AB^\dagger)^{L_2})(r=0)\rangle}.\\\nn
\eea
Following the same procedure of previous section, the above probability can be written at large $x$ limit as
\bea
\nonumber P\hspace{-.4in}&&\Big(\Tr((AB^\dagger )^J)^{L}(r=e^x)\rightarrow \Tr((AB^\dagger)^{L_1})(r=e^T)\Tr((AB^\dagger)^{L_2})(r=0)\Big)\\\nonumber
&=&\frac{\sum_{r_1\cdots r_L,s,s_1,s_2} g\Big([(R_1-r_1),1^{r_1}], \dots,[(R_L-r_L),1^{r_L}] ;[(S-s),1^{s}]\Big)^2\Big[f_{[(S-s),1^{s}]}^{N_1}f_{[(S-s),1^{s}]}^{N_2}\Big]^2}{ 2^{2LJ}J^{L}L!(N_1N_2)^{2LJ} L_1L_2}\\\nn &\times& g\Big([(S_1-s_1),1^{s_1}],[(S_2-s_2),1^{s_2}];[(S-s),1^s)] \Big)^2.
\eea
It is needless to say, one can compute other different types of correlators those have two final states like large number of gravitons creating two giant gravitons. However we close this section having these five above specific examples and in the next section we consider correlators involving more than two out going states.

\section{Higher genus factorization}
To get the large number of final states we should consider the higher genus $G= n-1$ factorization. Following \cite{Brown}, we also guess the probability for this condition as
\be
P(R \to R_1, R_2, \dots ,R_n)  =  \frac{1}{k_n^{2(\Delta_1 +\Delta_2 +\cdots + \Delta_n)}} \frac{ g ( R_1 , R_2, \dots, R_n ; R )^2 f_R^{N_1}f_R^{N_2}
}{ f_{R_1}^{N_1} f_{R_1}^{N_2} f_{R_2}^{N_1} f_{R_2}^{N_2} \cdots  f_{R_n}^{N_1}f_{R_n}^{N_2}}.
\ee
Here $k_n$ is a constant and it takes value 1 for genus zero and 2 for genus one. We are again computing the probability at long-distance limit, that is the operators are in a symmetric configuration far apart from each other.\\

\noindent
We find the probability for AdS giant with angular momentum $N_1$ going to $n$ number of smaller giants is,
\bea
&& P\Big([N_1]\rightarrow n\times[N_1/n]\Big) \nn \\
&& = \frac{1}{{k_n}^{2N_1}}\frac{f_{[N_1]}^{N_1}f_{[N_1]}^{N_2}}{{[f_{[\frac{N_1}{n}]}^{N_1}]}^n{[f_{[\frac{N_1}{n}]}^{N_2}]}^n} \nn \\
&& \hspace{-1 cm}\sim \frac{1}{\sqrt{2}}{\Big(\frac{n+1}{n}\Big)}^{\frac{n}{2}}\Big[\frac{4n^{n+1}}{k_n^2{(n+1)}^{n+1}}\Big]^{N_1}
{\Big(1+\frac{N_1}{N_2}\Big)}^{N_1+N_2-\frac{1}{2}} {\Big(1+\frac{N_1}{nN_2}\Big)}^{\frac{n}{2}-nN_2-N_1}.
\label{adshigher}
\eea
As a special case, we explicitly calculate the higher genus amplitude for $3$ outgoing smaller  AdS giants,
\bea
 &&P\Big([N_1]\rightarrow 3 \times[N_1/3]\Big)\nn\\
&=& \frac{e^{-\big[(3N_2+N_1-\frac{3}{2})\log(1+\frac{N_1}{3N_2})- (N_2+N_1-\frac{1}{2})\log(1+\frac{N_1}{N_2})+(6N_1-\frac{5}{2})\log(2)-(4N_1 -\frac{3}{2})\log(3)\big]}}{k_3^{2N_1}}.
\eea
In the same way the transition of a sphere giant to smaller giants is given by,
\bea
P\Big([1^{N_1}]\rightarrow n\times[1^{N_1/n}]\Big) \nn \\
= \frac{1}{{k_n}^{2N_1}}\frac{{f_{[1^{N_1}]}^{N_1}}{f_{[1^{N_1}]}^{N_2}}}{[f_{[1^{\frac{N_1}{n}}]}^{N_1}]^n[f_{[1^{\frac{N_1}
{n}}]}^{N_2}]^n} \nn \\
&& \hspace{-5 cm}\sim \frac{\sqrt{2\pi N_1}}{k_n^{2N_1}}{\Big(\frac{n-1}{n}\Big)}^{\frac{n}{2}+nN_1-N_1}{\Big(1-\frac{N_1}{N_2}\Big)}^{N_1-N_2-\frac{1}{2}}
{\Big(1-\frac{N_1}{nN_2}\Big)}^{\frac{n}{2} + n N_2-N_1}.
\label{spherehigher}
 \eea
Again in the case of sphere giant we calculate the transition amplitude for $3$ outgoing smaller giants.
\bea
& \hspace{-8 cm} P\Big([1^{N_1}]\rightarrow 3\times[1^{N_1/3}]\Big)  \nn \\
& = \frac{4}{k_3^{2N_1}}\sqrt{\frac{\pi}{27}}e^{-\big[(N_2-N_1+\frac{1}{2})
\log(1-\frac{N_1}{N_2})-(3N_2-N_1+\frac{3}{2})\log(1-\frac{N_1}{3N_2})-\frac{1}{2}\log(N_1)+2N_1(\log3-\log2)\big]}.
\eea
The transition of an $AdS$ giant carrying $\cal R$-charge $ \Delta_R $ into $n$ number of KK gravitons is given by,
\bea
&&P\Big([\Delta_R] \rightarrow \Tr((AB^\dag)^{\Delta_1})\cdots\Tr((AB^\dag)^{\Delta_n})\Big) \nn \\
&&= \frac{1}{k_n^{2\Delta_R}}
\frac{f_{[\Delta_R]}^{N_1} f_{[\Delta_R]}^{N_2} }{\langle\Tr((A^\dag B)^{\Delta_1})\Tr((AB^\dag)^{\Delta_1})\rangle\cdots\langle\Tr((A^\dag B)^{\Delta_n})\Tr((AB^\dag)^{\Delta_n})\rangle} \nn \\
&&\sim \frac{e^{-\big[2\Delta_R-(N_1+\Delta_R-\frac{1}{2})\log(1+\frac{\Delta_R}{N_1})
-(N_2+\Delta_R-\frac{1}{2})\log(1+\frac{\Delta_R}{N_2})\big]}}{k_n^{2\Delta_R}\Delta_1\cdots\Delta_n}.
\eea
The transition of an sphere giant carrying $\cal R$-charge $ \Delta_R $ into $n$ number of KK gravitons is given by,
\bea
&& P\Big([1^{\Delta_R}] \rightarrow \Tr((AB^\dag)^{\Delta_1})\cdots\Tr((AB^\dag)^{\Delta_n})\Big) \nn \\
&&= \frac{1}{k_n^{2\Delta_R}}
\frac{f_{[1^{\Delta_R}]}^{N_1} f_{[1^{\Delta_R}]}^{N_2} }{\langle\Tr((A^\dag B)^{\Delta_1})\Tr((AB^\dag)^{\Delta_1})\rangle\cdots\langle\Tr((A^\dag B)^{\Delta_n})\Tr((AB^\dag)^{\Delta_n})\rangle}\nn \\
&&\sim \frac{e^{-\big[2\Delta_R+(N_1-\Delta_R+\frac{1}{2})\log(1-\frac{\Delta_R}{N_1})
+(N_2-\Delta_R+\frac{1}{2})\log(1-\frac{\Delta_R}{N_2})\big]}}{k_n^{2\Delta_R}\Delta_1\cdots\Delta_n}.
\eea

\section{Transition probability in ABJM theory}
In this section we carry on our computation on transition probabilities among giant gravitons or from giant gravitons
to ordinary gravitons for ABJM theory where $N_1=N_2=N$. We will only enumerate the main results.\\

\noindent
The transition amplitude between two sphere giants of angular momentum $\frac{N}{2}$ to a single sphere giant of angular
 momentum $N$ is given
as
\be
P\Big(\Big[1^{\frac{N}{2}}\Big],\Big[1^{\frac{N}{2}}\Big] \rightarrow \Big[1^{N}\Big]\Big) = 2^{-2N}.
\ee
We extend the result for same type of transition occurring between AdS giants.
\be
P\Big(\Big[\frac{N}{2}\Big],\Big[\frac{N}{2}\Big] \rightarrow \Big[N\Big]\Big) = \frac{{\Big[(2N-1)!(N-2)!\Big]}^2}
{{\Big(\sum_{ i = 0}^{N_1/2 } (\frac{3N}{2}+i-1)!
(\frac{3N}{2}-i-2)!\Big)}^2}.
\ee
We calculate transition probability of the process depicting a sphere giant graviton is produced by the combination of $N$ number of KK gravitons of angular momentum 1.
\bea
&& P\Big(\Tr(AB^\dagger)^{N}(r=e^x)\rightarrow \chi_{[1^{N}]}(AB^\dagger)(r=0)\Big) \nn \\
&& \sim {\pi}^{\frac{1}{2}}e^{-\big[(N-\frac{1}{2})\log (N) + (N^2 +N - \frac{1}{2})\log(1 + \frac{1}{N}) -\frac{1}{2}\log(2)\big]}.
\eea
The transition probability to go from $N$ number of KK gravitons with angular momentum 1 to an AdS giant is
\bea
 P\Big(\Tr(AB^\dagger )^{N}(r=e^x)\rightarrow \chi_{[N]}(AB^\dagger)(r=0)\Big) \nn \\
\sim \pi^{-\frac{1}{2}} e^{-\big[N-\frac{1}{2N}+1+(N+\frac{1}{2})\log(N) -(4N-\frac{3}{2}) \log (2)\big]}.
\eea
The transition probability of the process where $L/J$ number of gravitons with angular momentum $J < \sqrt{N}$
 combining and giving a sphere giant of angular momentum $L$ is given by
\bea
P\Big(\Tr((AB^\dagger)^J)^{L/J}(r=e^x)\rightarrow \chi_{[1^{L}]}(AB^\dagger)(r=0)\Big) \nn \\
\sim \pi^{-\frac{1}{2}}e^{-\big[2L - \frac{L}{J}-\frac{1}{2}\log (J) +(\frac{L}{J}+\frac{1}{2})\log (L) + 2(N -L+\frac{1}{2}) \log (1- \frac{L}{N})+
\frac{1}{2} \log(2)\big]}.
\eea
Similarly for AdS giant the result changes as,
\bea
P\Big(\Tr((AB^\dagger)^J)^{L/J}(r=e^x)\rightarrow \chi_{[{L}]}(AB^\dagger)(r=0)\Big)  \nn  \\
\sim \pi^{-\frac{1}{2}}e^{-\big[2L-\frac{L}{J} - \frac{1}{2}\log (J) +(\frac{L}{J}+\frac{1}{2})\log (L) -2(N + L - \frac{1}{2}) \log (1 + \frac{L}{N})+
\frac{1}{2} \log(2)\big]}.
\eea
Now we calculate the transition probability from an AdS giant with angular momentum $N$ to two
smaller AdS giants with angular momentum $N/2$. The probability is
\be
\sim e^{-\big[2(3N-1)\log(3)-(8N-3)\log(2)]}.
\ee
For a sphere giant $[1^N]$ evolving into two smaller sphere giants $[1^{\frac{N}{2} }]$  probability
becomes
\be
\sim \pi e^{-\big[(4N+1)\log(2)-\log (N)\big]}.
\ee
The transition probability of evolving from one sphere giant with angular
momentum $N$ to two gravitons with angular momentum $N/2$ with the restriction $N/2$ is less than $\sqrt{N}$ is
\be
\sim \pi e^{-\big[2N+\log (N) + (2N-3)\log (2)\big]}.
\ee
We also calculate the transition probability of evolving from  one AdS giant with angular
momentum $N$ to two gravitons with angular momentum $N/2$ where we assume again the fact that  $N/2$ is less than $\sqrt{N}$ and the probability is
\be
\sim  e^{-2\big[N+\log (N) - (N+1/2)\log (2)\big]}.
\ee
Finally we calculate the higher genus transition probability for AdS giants. With the choice of $N_1 = N_2$,
(\ref{adshigher}) gives the appropriate higher genus correlator for the ABJM theory
\be
2^{2N - 1}{\Big(\frac{n+1}{n}\Big)}^{n-nN-N}{\Big[\frac{4n^{n+1}}{k_n^2{(n+1)}^{n+1}}\Big]}^N.
\ee
 Again for the case of sphere giants
equation (\ref{spherehigher}) modifies as,
\be
\frac{1}{k_n^{2N}}(2 \pi N)\Big(\frac{n-1}{n}\Big)^{n(1 + 2N)-2N}.
\ee
Having these discussion on transition probability in the last four sections we are going to find out the large $N$ expansion
of the theory in the non-trivial background.
\section{Large $N$ expansion in non-trivial background}
\noindent
We know that the large $N$ expansion of ${\cal N}=4$ SYM theory as well as of ABJM theory is replaced by $1/(N+M)$ in the
 presence of non-trivial background created by Young diagram of $N$ number of rows and $M$ number of columns of the order of $N$.
 Therefore for ABJ theory it is natural to expect the same. To verify that we compute the amplitude of the multi trace operator in
the non-trivial background as ABJM theory. Following \cite{Dmello,Koch4,dey,Dey}, we first calculate the amplitude without the
presence of background and the result is
\begin{eqnarray}
\nonumber{\cal{A}}\Big(\{n_i;m_j\},N_1,N_2\Big)&\equiv& \Bigg\langle\prod_{ij} \Tr\Big((AB^{\dagger})^{n_i}\Big
)\Tr\Big((A^{\dagger}B)^{m_j}\Big)\Bigg\rangle\nn\\
&=& \sum_{R,S}\alpha_R\beta_S\bigg\langle \chi_R(AB^{\dagger})\chi_S(A^{\dagger}B)\bigg\rangle \nn\\ &=&\sum_{R}\alpha_R\beta_R f_R^{N_1}f_R^{N_2}.
\label{corrwob}
\end{eqnarray}
Here we have rewritten the multi trace operator in terms of Schur polynomials as
\begin{equation}
\prod_i \Tr\Big((AB^{\dagger})^{n_i}\Big)= \sum_{R}\alpha_R \chi_R(AB^{\dagger}),\quad \quad\prod_j \Tr\Big((A^{\dagger}B)^{m_j}\Big)= \sum_{R}\beta_R \chi_R(A^{\dagger}B).
\end{equation}
Coefficients $\alpha_R$ and $\beta_R$ are independent of $N_1$ and $N_2$. The amplitude can also be calculated in the
presence of non-trivial background and the result is
\bea
{\cal{A}}_B\Big(\{n_i;m_j\},N_1,N_2\Big)&\equiv &\Bigg\langle\prod_{ij} \Tr\Big((AB^{\dagger})^{n_i}\Big)\Tr\Big((A^{\dagger}B)^{m_j}\Big)\Bigg\rangle_B\nn\\
&=&\sum_{R}\alpha_R\beta_R \Bigg(\frac{f_{+R}^{N_1}}{f_B^{N_1}}\Bigg) \Bigg(\frac{f_{+R}^{N_2}}{f_B^{N_2}}\Bigg).
\label{corrwb}
\eea
Here $f_B$ is the product of weights of the  background Young diagram $B$ and $f_{+R}$ is the product of the weights of the Young diagram $+R$ produced by the product of background Young diagram and Young diagram representing multi trace operator. All the weights of the diagram $B$ are repeated in the diagram $+R$.  Therefore all weights of $f_B$ will be canceled by the weights of the $f_{+R}$ and the remaining weights of the $f_{+R}$ contribute to find out the amplitude of the multi trace operator in presence of non-trivial background. However it seems that the remaining weights turn out the weights of the diagram $R$ which represent the multi trace operator, corresponding to the gauge group $U(N_1 + M)$ or $U(N_2 + M)$. Thus the amplitude with background can easily be calculated from the amplitude without background just by replacing the gauge group $U(N_1)$ and $U(N_2)$ as $U(N_1+M)$ and $U(N_2+M)$ respectively. Therefore we can write
\begin{equation}
{\cal{A}}_B\Big(\{n_i;m_j\},N_1,N_2\Big)={\cal{A}}\Big(\{n_i;m_j\},N_1+M,N_2+M\Big).
\label{compare}
\end{equation}
Since ${\cal{A}}\Big(\{n_i;m_j\},N_1,N_2\Big)$ admits expansions $1/N_1$ and $1/N_2$, so
 we can expect that ${\cal{A}}\Big(\{n_i;m_j\},N_1+M,N_2+M\Big)$ should have expansions $1/(N_1 + M)$ and $1/(N_2 + M)$.

\section{Conclusion:}
Half-BPS operators described by single trace operators of the dual theory of M/IIA theories with small $\cal R$-charge can be classified by Young tableaux. These operators can also be identified as the giant gravitons of the dual gravity theory. However, when the $\cal R$-charge of the gauge operators exceeds $N^{2/3}$, the role of
single trace operators is replaced by Schur polynomials. It is known that these polynomials play a very important role in studying the large $N$ expansion associated with the theory both for trivial and non-trivial backgrounds. In particular, for the later, when the background is provided by an operator with $\cal R$-charge of order $N^2$, the $1/N$ expansion rearranges itself to an expansion in $1/(N+M)$. Here $M$ is the number of columns in the representing Young diagram of the operator and it is of the order $N$. The purpose of this work has been to generalize these results for ABJ theory.\\

\noindent
In this paper, we have shown that, for  studying $U(N_1) \times U(N_2)$ (with say $N_1 < N_2$) ABJ theory, the $1/2$ BPS operators described by single trace operators, for $\cal R$-charge greater than  $\sqrt N_1$, are not appropriate. Rather, one needs to consider the Schur operators. We then identified the corresponding states in the dual gravity theory. Subsequently, we computed two, three and multi-point correlation functions
involving these operators. Our computations show, for large $N_1$ and $N_2$, all correlators with proper normalization converge to values less than unity -- a fact that is consistent with the probability interpretation of the correlators. We also have seen that two point correlators show a stringy exclusion principle.  While calculating the correlators, we used the overlap state and multi-particle normalization; both were found to depend on the gauge indices as well as the space-time coordinates. For the space-time part, the CFT factorization comes in to the picture. This, in turn, determines as to whether we choose the overlap state or the multi-particle state normalization for the outgoing states. We further found that the correlators of the gravity state have an exponential decay. However, owing to the parity non-invariance of the ABJ theory, we saw that the results are not symmetric under the exchange of $N_1$ and $N_2$. Finally, we considered a particular background produced by an operator with a $\cal R$-charge of ${\cal{O}}(N^2)$ and found that, in presence of this background, due to the non-planar contributions, the large $N_1$ and $N_2$ expansions get replaced by $1/(N_1 +M)$ and $1/(N_2 +M)$
respectively.\\

\noindent
{\bf Acknowledgements:} We would like to thank Robert de Mello Koch and Sudipta Mukherji for pleasant discussions, going through the draft and making valuable comments. The work of T.K.D. was supported by a DGAPA postdoctoral fellowship at the National Autonomous University of Mexico (UNAM), and by Mexico's National Council of Science and Technology (CONACyT) grant 104649.

\end{document}